\documentstyle[aps,preprint]{revtex}

\input epsf

\begin{document}

\preprint{\vbox{\hspace*{\fill} DOE/ER/40762-075\\
          \hspace*{\fill} U. of MD PP \#96-060}} \vspace{.5in}

\title{The High Temperature Phase of QCD and {\boldmath $U(1)_A$} Symmetry}

\author{Thomas D. Cohen}

\address{Department of 
Physics, University of~Maryland, College~Park, MD~20742}

\maketitle

\begin{abstract} 
\noindent Inequalities for QCD functional integrals are used to
establish that up to certain technical assumptions.
the high temperature chirally restored phase of QCD is effectively
symmetric under $U(N_f) \times U(N_f)$ rather than $SU(N_f) \times
SU(N_f)$.  If these assumptions are correct, there are no effects due
to anomalous breaking of $U(1)_A$ 
on correlation functions in this phase.  
 \end{abstract} 

\newpage
One of the most important features of QCD is that it has an approximate
$SU(N_f) \times SU(N_f)$ chiral symmetry which is spontaneously broken.
 For the purposes of this letter, it will be assumed that the symmetry
is exact or, more precisely, that corrections due to finite quark
masses are small and can be handled via chiral perturbation theory.  
If one were to study QCD above some critical temperature this
symmetry will be restored.  
The nature of this restored phase is of more than purely theoretical
interest since ultrarelativistic heavy ion collisions  are expected to
lead to thermalized regions of space with a temperature above $T_c$. 
For simplicity in the present discussion it will be assumed that, $N_f=2$. 

 This letter addresses the question of
what can be learned about the chirally restored phase directly from 
QCD via purely analytic means.   If one makes certain technical
assumptions, one  can deduce nontrivial---and rather
surprising---things about the nature of this phase by exploiting
QCD inequality techniques similar to those used by Wiengarten, Vaffa
and Witten\cite{QCDIN} in studies of QCD at $T=0$.   

In particular it will be shown that  if a certain set of zero measure
does not afflict the functional integral then, above $T_c$, the phase  is
effectively symmetric under $U(2) \times U(2)$ in the sense that
operators can be classified into  multiplets associated with
representations of $U(2) \times U(2) $ and that  correlation functions
of operators in a given multiplet are identical. This means, for example,
that the two-point  correlation function in the $\pi$ channel is
degenerate with the correlation function  in the $\eta^\prime$ channel.
 This is surprising since the  $U(2) \times U(2)$ symmetry of the QCD
lagrangian is broken by the $U(1)_A$ anomaly to $SU(2) \times SU(2)$. 
Moreover,  the $U(1)_A$ anomaly is at the operator level and thus the
anomaly exists independent of temperature.  As stressed by `t Hooft
\cite{Hooft}, the anomaly may be thought of as providing a mechanism
for explicit (as opposed to spontaneous) symmetry breaking.  Thus, it
seems {\it a priori} implausible
that restoration of chiral $SU(2) \times SU(2)$ should imply
invariance under $U(2) \times U(2)$.  Indeed, in a classic early review
of the subject of instantons and the $U(1)_A$ problem, Coleman\cite{Coleman} 
asserts precisely the viewpoint that the $U(1)_A$ symmetry remains
broken above the restoration temperature with the symmetry breaking
decreasing at high temperatures as a power of $1/T$.  Moreover
Meggiolaro \cite{Meggiolaro} has recently constructed a model motivated
by lattice calculations of the topological susceptability\cite{top} in
which $U(1)_A$ remains broken above the chiral restoration temperature.

The idea that   the chirally restored phase is  symmetric  under $U(2)
\times U(2)$  rather than  $SU(2) \times SU(2)$ is not new.  Shuryak 
raised this possibility previously\cite{Shuryak}.   The present
approach is novel, however, in that it derives this result formally from
the QCD functional integral.
 Before discussing the present derivation,   it is useful to review 
Shuryak's arguments.    One argument is based  on lattice calculations
of screening masses which purport to show that above $T_c$, the $\pi$
and $\sigma$ screening masses are degenerate (within numerical
noise)\cite{lattice}. The calculated  ``$\sigma$'' correlation
functions only included the quark-line connected part.   However, this
quark-line connected part is   the entire correlator in the
scalar-isovector ($\delta$) channel.  Thus, the lattice calculations
indicate a degeneracy  between the $\pi$ and $\delta$ screening masses.
 The $\pi$ and the $\delta$ do not belong  to the same  $SU(2) \times
SU(2)$ multiplet;  they do, however, belong to the same $U(2) \times
U(2)$ multiplet.  

Shuryak also argues for $U(2) \times U(2)$ restoration from  the
instanton liquid model\cite{ilm1,ilm2}.   Recall that the solution of
the $U(1)_A$ problem requires both the  $U(1)_A$ anomaly and the
contribution of nontrivial topological configurations\cite{Coleman}; 
topology is necessary for the anomalous violation of $U(1)_A$ symmetry
to have physical manifestations.  In the instanton liquid
model\cite{ilm1,ilm2} instantons provide the only source for these
configurations.  As has long been known, a finite density of 
instantons leads to chiral symmetry breaking\cite{cb}. In the instanton
liquid model, instantons also provide the only source of $SU(2) \times
SU(2)$ chiral symmetry breaking.
Thus, in this model, the same mechanism is responsible  both for
$SU(2) \times SU(2)$ chiral symmetry breaking and for allowing  the
$U(1)_A$ symmetry breaking due to the anomaly to have physical
consequences.  In such a model, if one were in a phase  in which $SU(2)
\times SU(2)$ chiral symmetry were unbroken, it would follow that
instanton effects are be turned off.  This, in turn,
suggests that the anomaly will not have physical effects in any of the
correlation functions: all observables will behave as though the phase
is  $U(2) \times U(2)$ symmetric.  
The mechanism responsible for this in the model is believed to be
the condensation of instantons and anti-instantons into topologically
neutral ``molecules''\cite{mol}.

The  present analysis  is based on   properties of
the QCD functional integral.  The essential physics is best understood
from the quark propagator in a given gluon background field.  When the
propagator is written in terms of a spectral representation, all
$U(1)_A$ violating effects come from eigenmodes in the neighborhood of
zero virtuality; {\it i.e.} $\lambda =0 $ modes, where the Dirac eigen
equation is $D\!\!\!\!\slash \psi_{j} = i \lambda_j \psi_{j}$.
  While it is not immediately obvious how to establish in general that
all  $U(1)_A$ violating amplitudes come from the region of $\lambda=0$,
it is  easy to establish for given $U(1)_A$ violating amplitudes by
studying the spectral representaton of the propagator  in the context
of the functional integral.  
Moreover,  it is easy to prove that  the density of states of the
Euclidean Dirac operator, $D\!\!\!\!\slash$ at $\lambda=0$, is zero for
{\it any} gauge configurations with boundary conditions consistent with
a temperature greater than the $T_c$ (excluding, perhaps, a set of zero
measure).  Thus, one can show that these $U(1)_A$ violating amplitudes vanish.

The fact that above $T_c$ all gauge configurations yield a vanishing density of
states at zero virtuality   can be seen quite transparently. The
chiral condensate $\langle \overline{q} q \rangle$
is related to the  density of states at zero averaged over gluon
field configurations\cite{CB}: $\langle \overline{q} q \rangle \, = \, -
\pi \langle \rho_A (\lambda = 0)  \rangle$
where $\rho_A$ is the density of states in a given background gluon
field configuration and $ \langle \, \, \,   \rangle$ indicates
averaging over the gluon field configurations weighted by ${\rm
e}^{-S_{\rm YM} }{\rm Det}[D\!\!\!\!\slash - m]$.  This applies to the
finite temperature case provided the average over gluons only includes
configurations periodic in Euclidean time with a periodicity of $\beta
=1/T$ and the fermion determinant is evaluated for antiperiodic
configurations.   Above $T_c$, 
$ \langle \langle \overline{q} q \rangle \rangle_T =0$, implying that
$\langle \langle \rho_A (0) \rangle \rangle_T=0$ (where the double
bracket indicates a thermal average).    However, $\rho_A(\lambda)$ is
a density; accordingly it is positive semi-definite ({\it i.e.}  $\ge
0$): $\rho_A (\lambda) \ge 0$.
Moreover the weighting function  ${\rm e}^{-S_{\rm YM}} {\rm
Det}[D\!\!\!\!\slash - m]$ is also positive semi-definite \cite{QCDIN}.
 An averaged  quantity which is never negative cannot have an average
of zero unless the quantity is zero for all configurations (except,
perhaps, a set of measure zero): $\rho_A(0) =0$ for configurations
consistent with the boundary conditions for $T > T_c$.

Before discussing how this works out in specific cases, it is worth
stressing the generality of the result.  It depends only on the fact
that $\rho_A(0)$ goes to zero above the phase transition for all
configurations and that $U(1)_A$ violating amplitudes come from modes
in the neighborhood of $\lambda=0$.   It does not depend on the
detailed mechanism which generates a nonzero  $\rho_A(0)$ below $T_c$. 
  
To make the  discussion concrete, consider the two-point correlation
function of scalar and pseudoscalar quark bilinears.
There are four distinct operators: pseudoscalar-isovector, $i
\overline{q}  \gamma_5 \tau q$, (the $\pi$ channel); scalar-isoscalar,
$\overline{q} q$ ($\sigma$); pseudoscalar-isoscalar, $i \overline{q} 
\gamma_5  q$, ($\eta^\prime$); 
and scalar-isovector, $\overline{q} \tau q$  ($\delta$).    These
bilinears are denoted as $J_\pi$, $J_\sigma$, $J_{\eta^\prime}$ and
$J_\delta$.   The $\sigma$ and $\pi$ form a distinct $SU(2) \times
SU(2)$ multiplet from the $\delta$ and $\eta^\prime$.  Under $U(2)
\times U(2)$ however they are all part of a single  multiplet.  

The thermal two-point correlation function $\Pi(\bf x)$ of two
equal-time quark bilinear operators at fixed temperature, $T$, is defined by 
\begin{equation}
\Pi_J ({\bf x}) \equiv  \langle \langle J({\bf x}) J({\bf 0}) \rangle 
\rangle_T \, -  \,  \langle \langle J({\bf x}) \rangle  \rangle_T
\langle \langle J({\bf 0}) \rangle  \rangle_T
\end{equation}
where the double braces indicates thermal average and $J({\bf x}) =
\overline{q}(x) \Gamma q(x)$ and  $\Gamma$ is a matrix in Dirac and
flavor space.  One can write this as a Euclidean functional integral:
\begin{eqnarray}
\Pi_J  ({\bf x}) \, & = \, & - \frac{1}{Z}\int_T \, D[A] \, \,e^{-S_{\rm
YM} } \,  {\rm Det}[D\!\!\!\!\slash - m_q] \,  \\ [.09in]
&  & \times \biggl[ {\rm tr}[S_A({\bf x},{\bf 0}) \Gamma S_A({\bf
x},{\bf 0}) \Gamma] \, - {\rm tr}[S_A({\bf x},{\bf x}) \Gamma ] \, {\rm
tr}[S_A({\bf 0},{\bf
0}) \Gamma]  \biggr] \nonumber
\end{eqnarray}
where the subscript $T$ indicates the finite $T$ boundary conditions
(periodic in $A$); $Z$ is the partition function; $S_{\rm YM}$ is the
Euclidean action of the Yang-Mills field; the Det indicates a
functional determinant with the
fermion modes satisfying antiperiodic boundary conditions; $S_A({\bf
x}, {\bf y})$  is the Euclidean space quark propagator in the presence
of a background gauge field $A$; and the 
traces are over color, flavor and Dirac spaces.

A finite quark mass $m_q$ is included in the previous expression---it
will be sent to zero only at the  end of the calculation.    For
technical reasons it is simpler to work in a box of finite volume $V$
(which makes all of the modes discrete)  and to let the volume of the
box go to infinity at the end of the problem.  The ordering of these
two limits is critical.  One must take the $V \rightarrow \infty$ limit
before taking the chiral limit.\cite{LS}  

There are two distinct contributions to this functional integral: a
term with a single trace and a term with two traces.  They are the
quark-line connected and quark-line disconnected pieces, respectively.
If the up and down quark masses  are equal (as is assumed here),  
$[S_A({\bf x},{\bf 0}), \tau]  =0$, and the connected piece of an
isoscalar correlator ({\it e.g.} the $\sigma$ channel) is identical to
the connected piece of an isovector correlator with the same spatial
quantum numbers ({\it e.g. }the $\delta$ channel). 

The difference between  the $\sigma$ and $\delta$ correlation
functions, $\Pi_\sigma (x) - \Pi_\delta(x)$, is $U(1)_A$ violating.  As
noted above, in the functional integral this difference comes entirely
from the quark-line disconnected piece, 
\begin{equation}
\Pi_\sigma({\bf x}) \, - \, \Pi_\delta({\bf x}) \, = \label{c1}\\
\, \frac{1}{Z} \int_T \, D[A] \, e^{-S_{\rm YM} }\, {\rm Det}
[D\!\!\!\!\slash - m_q] \,  {\rm tr}[S_A({\bf x},{\bf x})  ] \, {\rm
tr}[S_A({\bf 0},{\bf 0}) ] \; \; \; .  \nonumber
\end{equation}
If it can be shown that above the  $SU(2) \times SU(2)$ chiral
restoration temperature
\begin{equation} 
{\rm tr}[S_A({\bf x},{\bf x})  ] \sim {\cal O}(m_q) \label{van}
\end{equation}
for {\it all} gauge configurations consistent with the boundary
conditions, then it follows that  ${\rm tr}[S_A({\bf x},{\bf x})  ]
{\rm tr}[S_A({\bf 0},{\bf 0}) ] \sim {\cal O}(m_q^2)$ for all gauge
configurations and thus the weighted average over gauge configurations
will also be $  {\cal O}(m_q^2)$ from which eq.~(\ref{c1}) implies
\begin{equation}
\Pi_\sigma({\bf x}) \, - \, \Pi_\delta({\bf x}) \, \sim {\cal
O}(m_q^2)\; \; \; . \label{u11}
\end{equation}
This in turn implies that in the chiral limit of $m_q \rightarrow 0$
$\Pi_\sigma({\bf x}) \, - \, \Pi_\delta({\bf x}) \, \rightarrow 0$. 
That is, this $U(1)_A$ violating matrix element vanishes.

If the validity of eq.~(\ref{van}) is established, then one has proven
that this $U(1)_A$ violating amplitude vanishes.   To begin use a
spectral representation for $S_A$:
\begin{equation}
S_A({\bf x},{\bf y}) = \sum_j \frac{\psi_j ({\bf x}) \, \psi_j^\dagger
({\bf y}) }{i \lambda_j - m_q}
\end{equation}
where the modes are eigenmodes of the Dirac operator.
>From the fact that $\{\gamma_5,D\!\!\!\!\slash\} =0$, it follows that
if $\psi_j$ is an eigenmode with eigenvalue $i \lambda_j$, then
$\gamma_5 \psi_j$ is an eigenmode with eigenvalue $-i \lambda_j$.
This in turn implies
\begin{equation} 
{\rm tr}[ S_A({\bf x},{\bf x}) ] = \sum_j \frac{- m_q \psi_j^\dagger
({\bf x}) \, \psi_j ({\bf x}) }{\lambda_j^2 + m_q^2} \; \; \; .  \label{spec1}
\end{equation}
   It is  apparent from eq~(\ref{spec1}) that, as advertised, in the
limit of  $m_q \rightarrow 0$, contributions to ${\rm tr}[ S_A({\bf
x},{\bf x}) ] $ come entirely from  modes near $\lambda = 0$.  More
significantly, given the standard convention that the quark mass is
positive, then ${\rm tr}[ S_A({\bf x},{\bf x}) ] \le 0$ for any  gauge
configuration.
 
 Above $T_c$, we know that 
chiral condensate, $\langle \langle \overline{q} q \rangle \rangle_T$,
vanishes, or to be more precise is order $m_q$ and vanishes when the 
chiral limit is taken.  The chiral condensate can be written as a
functional integral:
\begin{equation}
N_f \langle \langle \overline{q} q({\bf x}) \rangle \rangle_T\,  =  \label{cc}
\\  \frac{1}{Z}\int_T \, D[A] \, \,e^{-S_{\rm YM} } \,  {\rm
Det}[D\!\!\!\!\slash - m_q]  \, {\rm tr}[S_A({\bf x},{\bf x})] \, \sim 
- {\cal O}(m_q) \nonumber
\end{equation}
At this stage, it is worth recalling that  $e^{-S_{\rm YM} } {\rm
Det}[D\!\!\!\!\slash - m_q] $ is positive semi-definite for all gauge
configurations while $ {\rm tr}[S_A({\bf x},{\bf x})] $
is negative semi-definite.  Thus, the integrand in eq.~(\ref{cc}) is
negative  semi-definite.   This means that there can be no
cancellations in the integral---the only way that the integral can be 
${\cal O}(m_q)$ is if the contributions from all gauge configurations 
are ${\cal O}(m_q)$ (except perhaps from a fraction of  configurations
which goes to zero in the chiral limit).   Thus eq.~(\ref{van}) has
been shown to be true for  all gauge configurations contributing to the
functional integral except for contributions which become a set of
measure zero in the chiral limit.  Assuming this set of measure zero
can be safely ignored
in the evaluation of eq.~(\ref{c1})---an issue which will be discussed
at the end of this letter---one concludes that since eq.~(\ref{van}) is
true so is  eq.~(\ref{u11}); thus in the chiral limit of $m_q
\rightarrow 0$ these $\sigma$ and $\delta$ correlators are identical.

Having established this, it is immediately obvious that $\Pi_\pi ({\bf
x})$, $\Pi_{\eta^\prime} ({\bf x})$, $\Pi_\sigma ({\bf x})$, and
$\Pi_\delta ({\bf x})$ must all be identical above $T_c$ in the $m_q
\rightarrow 0$ limit.  $SU(2) \times SU(2)$ chiral restoration implies
that  $\Pi_\pi ({\bf x})=\Pi_\sigma ({\bf x})$ and $\Pi_{\eta^\prime}
({\bf x})=\Pi_\delta ({\bf x})$, while eq.~(\ref{u11}) implies that
$\Pi_\delta ({\bf x})=\Pi_\sigma ({\bf x})$ --- all members of the $U(2)
\times U(2)$ multiplet are identical.  Although from this argument it
is clear that $\Pi_{\eta^\prime} ({\bf x}) = \Pi_\pi ({\bf x})$, it is
useful to demonstrate this directly from functional integral
inequalities as it demonstrates a technique which is useful for
studying other multiplets.  

The functional integral for this difference can be written as 
\begin{equation}
\Pi_\pi ({\bf x}) \, - \, \Pi_{\eta^\prime}({\bf x}) \,  =
\label{c2}\\  \frac{1}{Z} \int_T \, D[A] \, e^{-S_{\rm YM} }\, {\rm
Det} [D\!\!\!\!\slash - m_q] \, {\rm tr}[S_A({\bf x},{\bf x}) 
\gamma_5] {\rm tr}[S_A({\bf 0},{\bf 0}) \gamma_5] \: .\nonumber
\end{equation}
The first step in proving that $\Pi_\pi({\bf x}) \, - \,
\Pi_{\eta^\prime}({\bf x})$ goes to zero above $T_c$ is to show that 
\begin{equation} 
\left | {\rm tr} [S_A({\bf x},{\bf x})  \gamma_5] \right | \le \left |
{\rm tr} [S_A({\bf x},{\bf x}) ] \right |
\label{i1}
\end{equation}
 for any gauge configurations.
This is easily established using the spectral decomposition  of the
propagator and the fact that $\psi_j^\dagger ({\bf x}) (1 + \gamma_5)^2
 \psi_j(\bf{x}) \ge 0$ for any $\psi_j$.
 By comparing eq.~(\ref{c1}) with eq.~(\ref{c2}) and using 
eq.~(\ref{i1}) and the fact that $e^{-S_{YM}} {\rm Det}[D\!\!\!\!\slash
- m_q]$ is positive semi-definite, one sees  that 
$\left | \Pi_\pi({\bf x}) \, - \, \Pi_{\eta^\prime}({\bf x}) \right |
\le \left | \Pi_\sigma ({\bf x}) \, - \, \Pi_\delta({\bf x}) \right | $. 
Since the right-hand side of this inequality goes to zero,  the
left-hand side does as well, and thus the degeneracy of the $\pi$ and
$\eta^\prime$ channels above $T_c$ has been demonstrated directly from
the QCD functional integrals.

The technique used to establish that $\pi$ and $\eta^\prime$
correlation functions are  identical above the phase transition by
showing that the absolute value of their difference is less than or
equal to  $\left| \Pi_\sigma -\Pi_\delta \right |$, can be immediately
generalized for other channels.
In this way, one can show that vector and pseudo-vector, isovector and
isoscalar ({\it i.e.} the $\omega$, $\rho$, $f_1$ and $a_1$) correlation functions 
are all identical above $T_c$.   Again this is an identification of
$U(2) \times U(2)$ symmetry since only the $\rho$ and $a_1$ are
connected by $SU(2)\times SU(2)$ chiral symmetry.
The same method allows one to show that the  tensor and pseudo-tensor,
isoscalar and isovector correlators
are identical above $T_c$.

There is a loophole in the demonstration of the $U(2) \times U(2)$
nature of the chirally restored phase given above.  In particular, it
was assumed that contributions to the functional integral in
eq.~(\ref{cc}), which were a set of measure zero in the $m_q \rightarrow
0$ limit, do not  contribute to the functional integral 
in eq.~(\ref{c1}).   By inspection, it is clear that so long as ${\rm
tr}[S_A({\bf x},{\bf x})]$ is finite for all gauge configurations (after
a gauge invariant and $SU(2) \times SU(2)$ chiral invariant ultraviolet
regularization), then the set of measure zero cannot affect 
the functional integral in eq.~(\ref{c1}).  To discuss a set of measure
zero with infinite contributions it is sensible to first introduce an
infrared cutoff regulator, $\epsilon$ where $\epsilon \rightarrow 0$
corresponds to the infinite volume and $m_q \rightarrow 0$ limits with 
$V m_q^3  \rightarrow \infty$.   The loophole in the general argument
is  that there could be configurations for which 
${\rm tr}[S_A({\bf x},{\bf x})] \sim \epsilon^{-1/2}$ which have a
weight proportional to $\epsilon$.  In such a case, $\langle \langle
\overline{q}q({\bf x}) \rangle \rangle_T \sim \epsilon^{1/2}$ which
vanishes in the $\epsilon \rightarrow 0$ limit while $\{\Pi_\sigma({\bf
x}) - \Pi_\delta({\bf x}) \} \sim {\cal O}(1)$.  Thus, there is
apparently the possibility that the chiral condensate vanishes as the
regulator goes to zero  while the $U(1)_A$ violating amplitude,
$\{\Pi_\sigma({\bf x}) - \Pi_\delta({\bf x}) \}$, does not.

In summary, up to the loophole discussed above, it has been shown
directly from the QCD function integral
that above $T_c$,  the correlation functions for quark bilinears in a
given $U(2) \times U(2)$ multiplet are identical.   This indicates that
the phase is invariant under $U(2) \times U(2)$ rather than $SU(2)
\times SU(2)$.    The anomalous $U(1)_A$ breaking does not split the
$U(2) \times U(2)$ multiplets because the effects of the anomaly occur
entirely through the quark-line disconnected parts of correlation
functions and, in the $m_q \rightarrow 0$ limit, the quark-line
disconnected parts contribute only due to the the modes near
$\lambda=0$.  Above $T_c$ the density of states at $\lambda=0$ goes to
zero and the anomaly ceases to play a role.

 The author thanks V. Soni and W. Melnitchouk for useful discussions
and to Thomas Sch\"{a}fer for introducing him to the results of the
instanton liquid model.   The early stages of this work was done during a
visit to the Department of Physics and the Institute for Nuclear Theory
at the University of Washington; their kind hospitality is gratefully
acknowledged.  This work was supported in part by the U. S. Department
of Energy (grant
no.  DE-FG02-93ER-40762) and the U.  S. National Science Foundation
(grant no. PHY-9058487).

\end{document}